\documentclass[12pt]{article}
\pdfoutput=1
\usepackage[utf8]{inputenc}
\usepackage{amsmath}
\usepackage{amssymb}
\usepackage{units}
\usepackage{graphicx}
\usepackage{slashed}
\usepackage{array}
\usepackage[leftcaption,raggedright]{sidecap}
\usepackage{caption}
\usepackage[hidelinks]{hyperref}
\usepackage[inkscapelatex=false]{svg}

\newcommand\pubnumber{NuPhys2023-Pablo-Barham-Alzás}
\newcommand\pubdate{\today}

\def\napoli{European Organisation for Nuclear Research (CERN)}

\def\support{\footnote{
  pablo.barham@cern.ch
}}

\def\Title#1{\begin{center} {\Large #1 } \end{center}}
\def\Author#1{\begin{center}{ \sc #1} \end{center}}
\def\Address#1{\begin{center}{ \it #1} \end{center}}

\newcommand\pubblock{\rightline{\begin{tabular}{l} \pubnumber\\
         \pubdate  \end{tabular}}}
\newenvironment{Abstract}{\begin{quotation}  }{\end{quotation}}
\newenvironment{Presented}{\begin{quotation} \begin{center} 
             PRESENTED AT\end{center}\bigskip 
      \begin{center}\begin{large}}{\end{large}\end{center} \end{quotation}}

\def\beq{\begin{equation}}
\def\eeq#1{\label{#1}\end{equation}}
\def\eeqn{\end{equation}}

\def\beqa{\begin{eqnarray}}
\def\eeqa#1{\label{#1}\end{eqnarray}}
\def\eeqan{\end{eqnarray}}

\let\bar=\overbar

\def\Dslash{\not{\hbox{\kern-4pt $D$}}}
\def\dslash{\not{\hbox{\kern-2pt $\del$}}}

\def\msb{{\bar{\ssstyle M \kern -1pt S}}}

\begin{document}
\begin{titlepage}
\pubblock

\vfill
\Title{}
\vfill
\Author{Pablo Barham Alzás\support, Radi Radev}
\Address{\napoli}
\vfill
\begin{Abstract}
\textbf{Differentiable nuclear deexcitation simulation for low energy neutrino physics.}

Neutrino-nucleus interactions play an important role in present and future neutrino experiments. The accurate simulation of these interactions at low energies ($<$100 MeV) is crucial for the detection and study of supernova, solar and atmospheric neutrinos. In particular, the reconstruction of the incoming neutrino properties depends on the ability to measure the products from the deexcitation of the final state nucleus after the initial neutrino-nucleus scattering reaction. A realistic nuclear deexcitation model that can correctly manage the theoretical uncertainties in the process is key to determine the response of a detector to low energy neutrinos and estimate the overall systematic uncertainties that affect it.

Automatic differentiation frameworks like PyTorch or JAX, widely used in the field of Machine Learning, provide us the tools to compute exact gradients of the simulation outputs with respect to the model parameters. Such differentiable simulators can be applied in simulator tuning to match observed data and forward modeling to efficiently infer the impact of parameter distributions to the physics output, or for parameter inference acting as a likelihood estimator. This paradigm can pave the way to a fully differentiable analysis of the whole simulation chain or direct integration with preexisting Machine Learning tools.
As a proof of concept, in this work we implement a fully differentiable nuclear deexcitation simulation based on a simplified version of the Hauser-Feshbach statistical emission model.

\end{Abstract}
\vfill
\begin{Presented}
NuPhys2023, Prospects in Neutrino Physics\\
King's College, London, UK,\\ December 18--20, 2023
\end{Presented}
\vfill
\end{titlepage}
\def\thefootnote{\fnsymbol{footnote}}
\setcounter{footnote}{0}

\section{Introduction}
The modelling of neutrino-nucleus interactions is key in the development present and future neutrino experiments like DUNE. Accurate simulation of these interactions at low energies ($<$100 MeV) is crucial for the detection and study of supernova and solar neutrinos \cite{bib:low-e-lartpcs}. At these energies, it is a common approach to factorise the neutrino-nucleus interaction in two parts:
\begin{enumerate}
    \item First, the initial lepton-nucleus scattering process is considered, leaving the final state nucleus in an excited state: $\nu + N \rightarrow e + \Tilde{N}^*$.
    \item Starting from this excited nuclear state, we can compute the probability of decaying to a lower lying state via the emission of a deexcitation gamma or nuclear fragment: $\Tilde{N}^* \rightarrow \Tilde{N} + \text{(deexcitation products)}$.
\end{enumerate}
These two steps are simulated in order and independently, meaning that the deexcitation process is independent of the lepton kinematics. However, the reconstruction of the incoming neutrino properties depends on the ability to measure the products coming from the this deexcitation process. A realistic nuclear deexcitation model that can manage the relevant theoretical uncertainties is therefore key to determine the response of a detector to low energy neutrinos.

\section{Hauser-Feshbach nuclear deexcitation model}
The Hauser-Feshbach model \cite{bib:h-f} is a statistical emission model based on the assumption that a neutrino-nucleus scattering at low energies is essentially a \textit{compound-nucleus} reaction: the energy is rescattered among all the particles of the final state nucleus, reaching thermal equilibrium, and deexcitation to a lower energy state can then be regarded as a competition between all the available decay mechanisms: emission of a nuclear fragment (proton, neutron, deuteron, alpha particle...) or a photon. This assumption can be justified by looking at electron-nucleus scattering data for similar energies \cite{bib:escat}.
We follow the MARLEY event generator \cite{bib:marley} approach in which we can write the differential decay width to a given state by emitting a particular nuclear fragment $a$ as
\begin{equation}
    \frac{d \Gamma_a}{d E_f}=\frac{1}{2 \pi \rho_i\left(E_i, \alpha\right)} \sum_{\alpha^{\prime}} T_{\alpha^{\prime}}(\varepsilon) \rho_f\left(E_f, \alpha^{\prime}\right).
\end{equation}
Above, $\rho_i$ and $\rho_f$ are the event densities at the initial and final nuclear states, $T$ is the transition strength function, and the sums are performed over the different quantum numbers $\alpha$ involved in the process: EM transition type and multipolarity for $\gamma$-rays or angular momentum for massive fragments, and isospin and total angular momenta of the intial and final nuclear states.

\section{Differentiable simulation implementation}
Automatic differentiation frameworks like JAX \cite{bib:JAX} provide us the tools to build differentiable simulators by computing exact gradients of the simulation outputs with respect to the model parameters. Making a simulation differentiable offers distinct advantages like:
\begin{itemize}
    \item Model parameter tuning via gradient descent.
    \item Efficient estimation of the impact of parameter distributions on the physical observables.
    \item Integration with existing or bespoke Machine Learning tools.
\end{itemize}
We implement a differentiable toy version of the Hauser-Feshbach nuclear deexcitation model, where we consider a single nucleus type and only include the dominant mode of deexcitation: $\gamma$-ray emission.

\subsection{Sampling the continuum}
Starting with an initial nuclear energy $E_0$ transferred by the neutrino in an interaction, we sample the next excitation energy via the inverse CDF sampling path:
\begin{equation}
    \hat{E}_1 = g(\hat{\epsilon}; E_0, \boldsymbol{\theta}); \quad \hat{\epsilon} \sim \mathcal{U}[0,1].
\end{equation}
Here $\boldsymbol{\theta}$ are the model parameters, $g$ is the inverse CDF of the differential decay width modified to account for the probability of the decay to a discrete nuclear state occurring, and $\hat{\epsilon}$ is sampled from the uniform $[0,1]$ distribution. We continue sampling energies sequentially: $\hat{E}_{n+1} = g(\hat{\epsilon}; E_n, \boldsymbol{\theta})$, until we reach a bound (discrete) state.

\subsection{Estimating the gradients}
To estimate the gradients of the expected values of a physical observable $f$ (``stochastic gradients") with respect to the model parameters $\boldsymbol{\theta}$, we use a pathwise gradient estimator, also known as the ``reparameterisation trick" \cite{bib:mge}:
\begin{equation}
    \nabla_{\boldsymbol{\theta}} \mathbb{E}_{p\left(E_i ; E_{i-1}, \boldsymbol{\theta}\right)}\left[f\left(E_i\right)\right]=\mathbb{E}_{p\left(E_i ; E_{i-1}, \boldsymbol{\theta}\right)}\left[\nabla_{E_i} f\left(E_i\right) \nabla_{\boldsymbol{\theta}} E_i\right].
\end{equation}
Computing the second term in this expression, $\nabla_{\boldsymbol{\theta}}E_i$ involves in principle taking the derivative of the inverse cumulative distribution function (CDF) $g$. As analytically inverting an arbitrary function is hard, we can rewrite this term using implicit differentiation as
\begin{equation}
    \nabla_{\boldsymbol{\theta}} E_i=-\left(\nabla_{E_i} g^{-1}\left(E_i ; E_{i-1}, \boldsymbol{\theta}\right)\right)^{-1} \nabla_{\boldsymbol{\theta}} g^{-1}\left(E_i ; E_{i-1}, \boldsymbol{\theta}\right).
\end{equation}
This allows us to decouple the sampling procedure from the gradient computation and obtain the gradient directly from the CDF $g^{-1}$. Note that in the above equations we must be careful to include the dependence of the previous energy with the model parameters $\boldsymbol{\theta}$.

\subsection{Gradients of the discrete levels}
The deexcitation process continues sampling unbound (continuum) levels until a bound energy is reached. Once one of these discrete levels is hit, we no longer have a smooth dependence of the process energies on the model parameters. Taking derivatives of discretely distributed random variables is problematic, and although several approaches have been proposed for an object that behaves in a desired fashion (see e.g. \cite{bib:dd}), we sidestep the issue by instead sampling the whole ``discrete tree": we compute the probability of each discrete path and it's derivative with respect to the model parameters $\boldsymbol{\theta}$. Then, we can estimate the elements of the gradient as
\begin{equation}
    \left[E_i\right]=\sum_{\text {paths }} p(\text {path}) \cdot E_{i, \text {path}} \longrightarrow\left\{\begin{array}{l}
\nabla_{\left[E_i\right]} f\left(\left[E_i\right]\right)=\sum_{\text {paths }} p(\text {path}) \cdot \nabla_{E_i} f\left(E_{i, \text {path}}\right) \\
\nabla_{\boldsymbol{\theta}}\left[E_i\right]=\sum_{\text {paths }} \nabla_{\boldsymbol{\theta}} p(\text {path}) \cdot E_{i, \text {path}}
\end{array}\right.
\end{equation}
where $[E_i]$ is the expected value of the energy over the discrete paths.

\section{Results}
We have successfully implemented our differentiable toy model. Our sampling procedure is vectorised and runs of the GPU, with gradients checked against finite differences.
We can also estimate the uncertainty of the expected values of an observable given a set of parameters by running the simulation on the central values only and performing an expansion around them.

\begin{figure}[ht!]
    \centering
    % First figure
    \begin{minipage}[b]{0.49\linewidth}
        \centerline{\includegraphics[width=1\columnwidth]{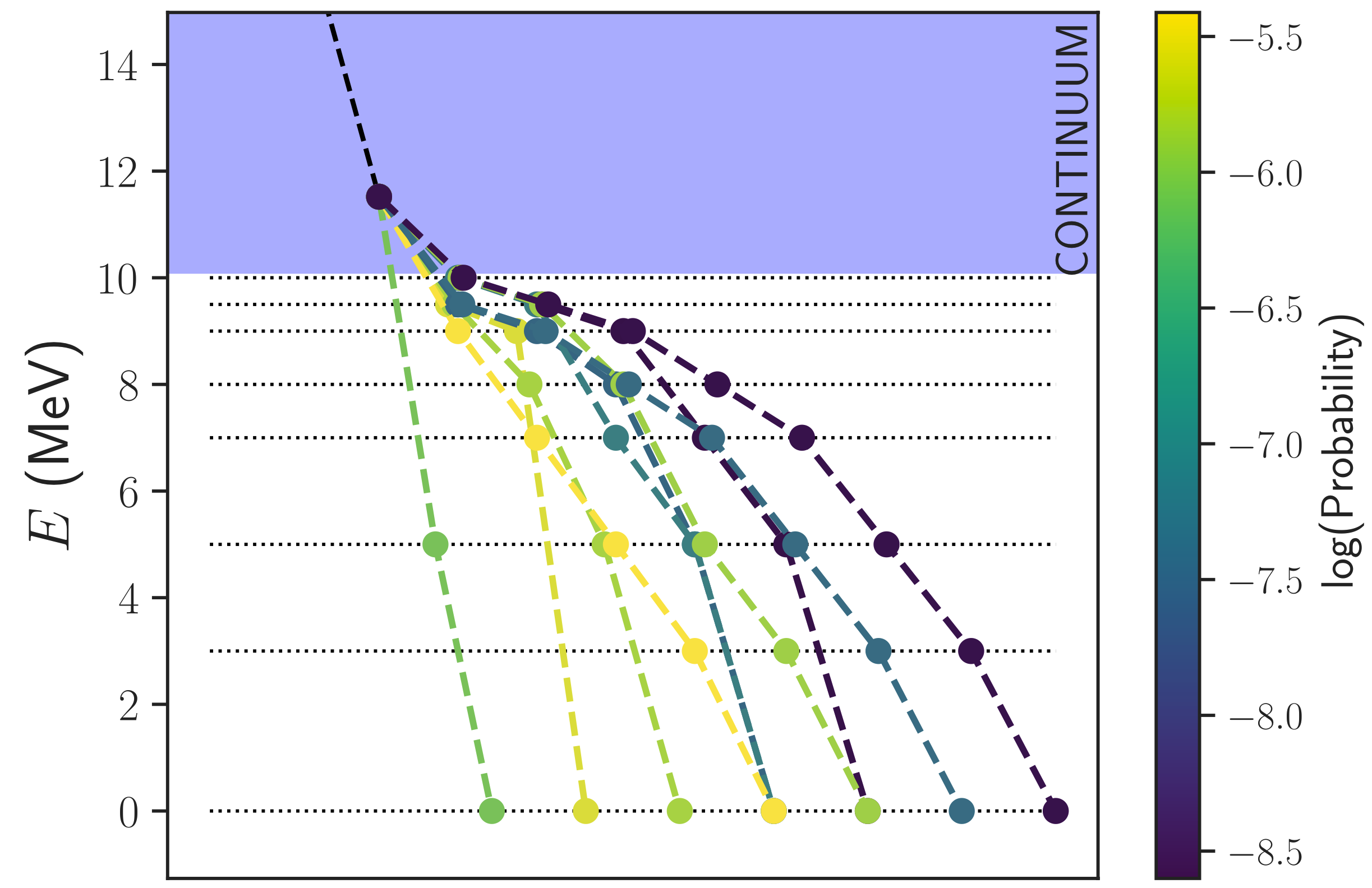}}
        \caption{Subsample of the ``discrete tree" coloured by path probability.}
        \label{fig:discrete_tree_prob}
    \end{minipage}
    \hfill % Adds horizontal space between the figures
    % Second figure
    \begin{minipage}[b]{0.49\linewidth}
        \centerline{\includegraphics[width=1\columnwidth]{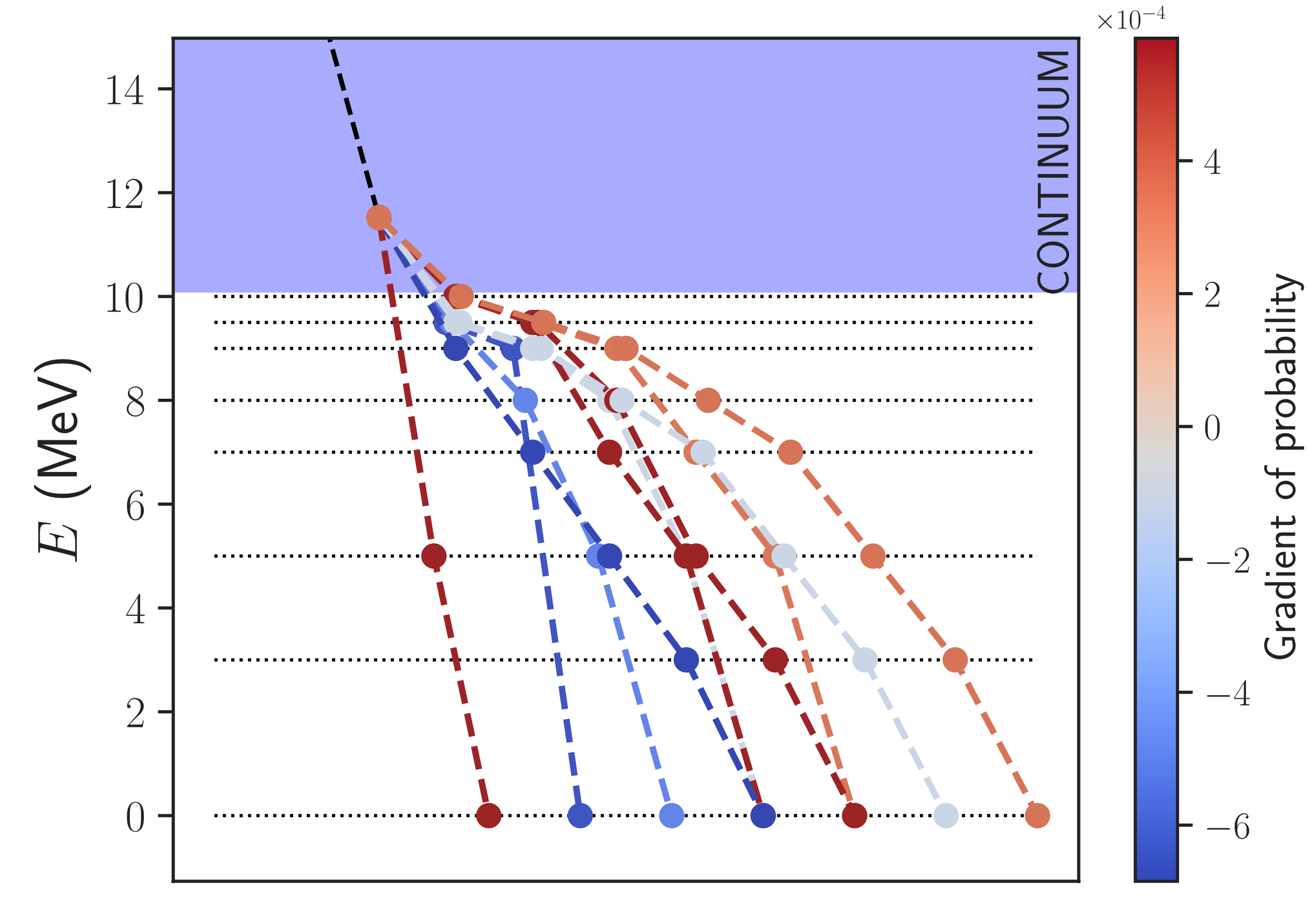}}
        \caption{Subsample of the ``discrete tree" coloured by path gradient w.r.t. $\alpha$.}
        \label{fig:discrete_tree_grad}
    \end{minipage}
\end{figure}

A subsample of the discrete ``deexcitation tree" is illustrated in Figure \ref{fig:discrete_tree_prob}, showing paths coloured by the path probability, and Figure \ref{fig:discrete_tree_grad}, where the paths are coloured by the gradient of the path probability with respect to model parameter $\alpha$.

\begin{figure}[ht!]
    \centering
    % First figure
    \begin{minipage}[b]{0.49\linewidth}
        \centerline{\includegraphics[width=1\columnwidth]{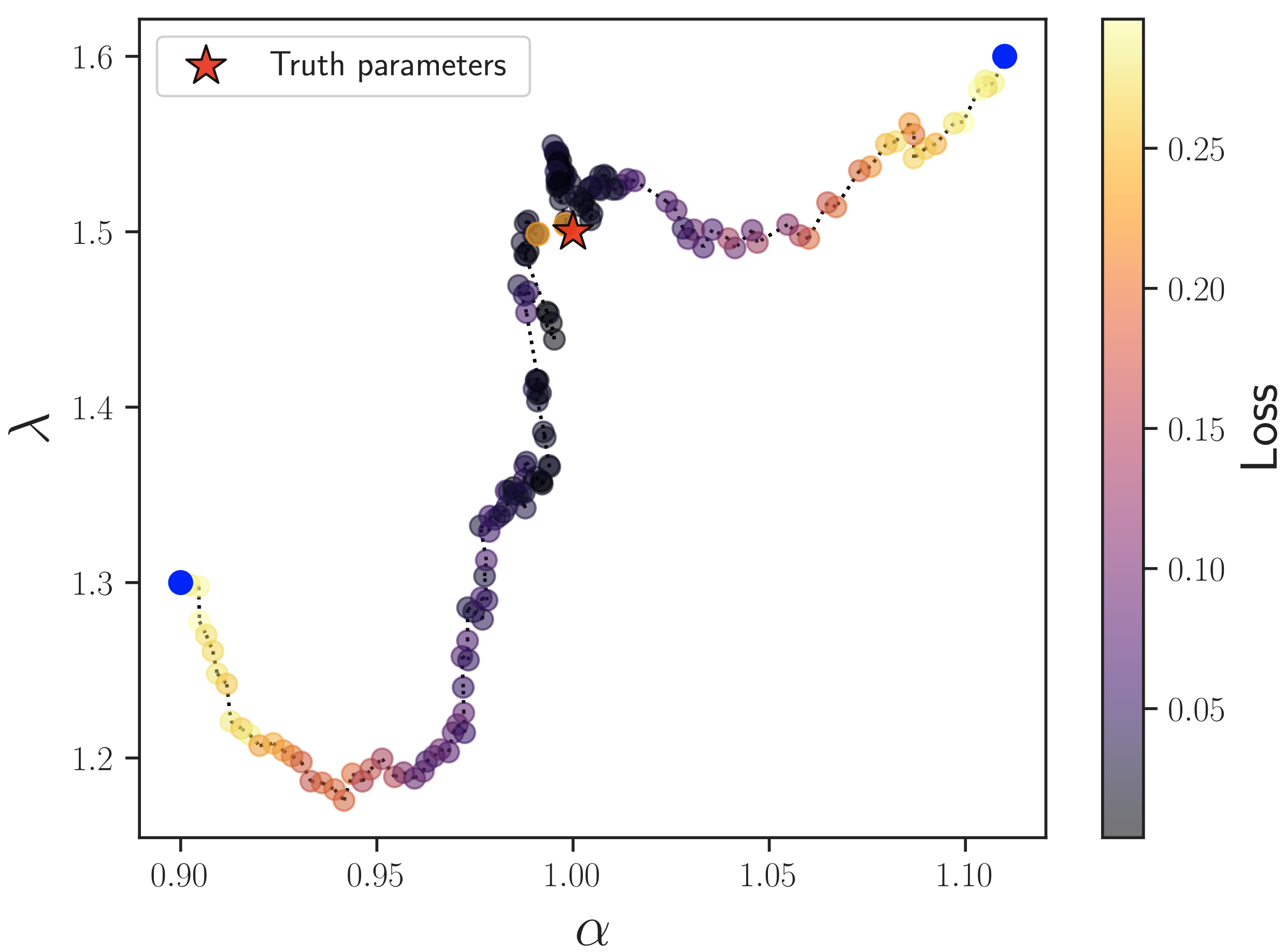}}
        \caption{Two gradient descent iterations converging in the $\alpha-\lambda$ plane, coloured by loss.}
        \label{fig:params_evolution}
    \end{minipage}
    \hfill % Adds horizontal space between the figures
    % Second figure
    \begin{minipage}[b]{0.49\linewidth}
        \centerline{\includegraphics[width=1\columnwidth]{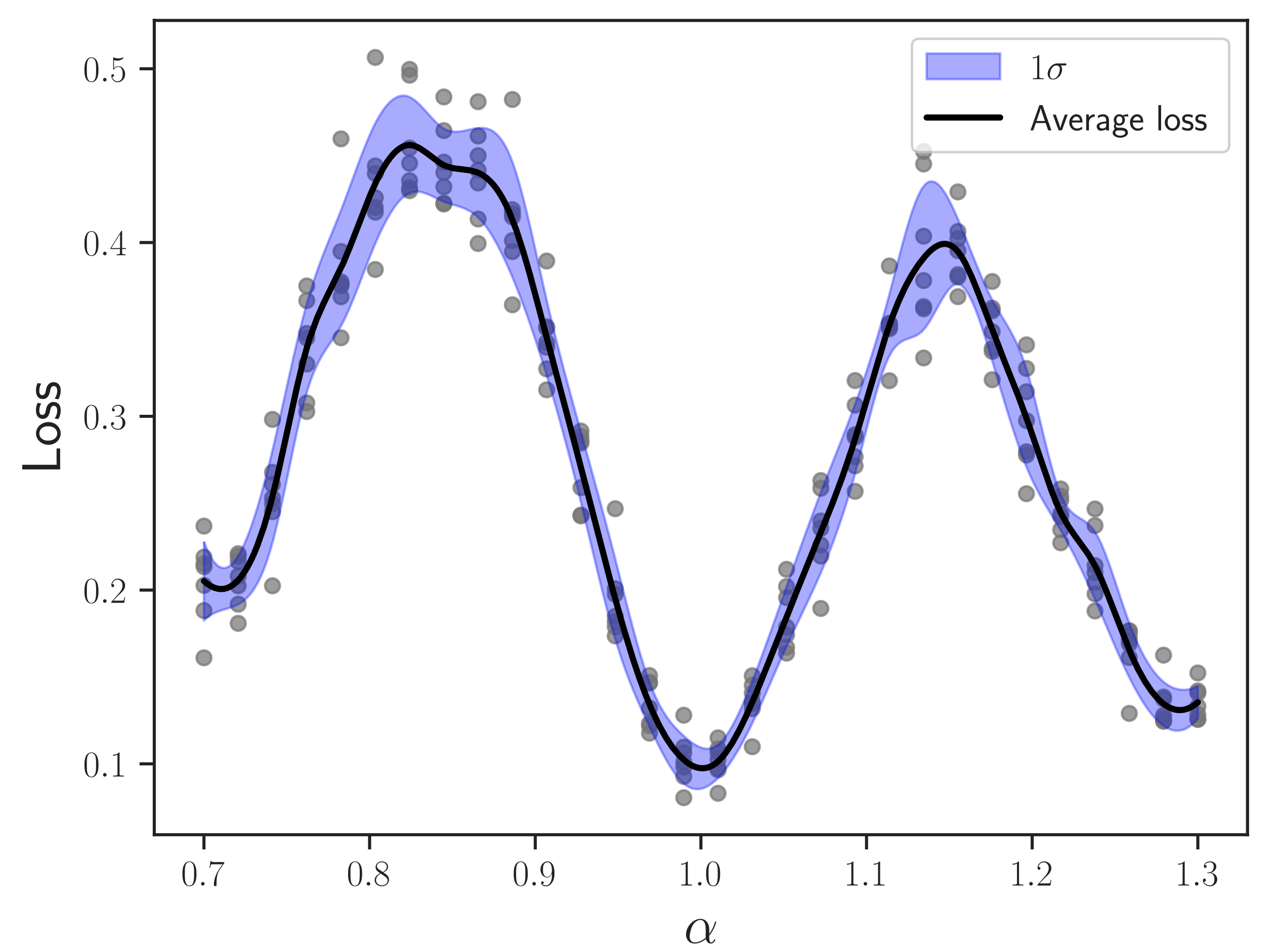}}
        \caption{Loss function for different values of the $\alpha$ parameter.}
        \label{fig:loss_alpha}
    \end{minipage}
\end{figure}

An immediate application of the differentiable simulation is model fitting via gradient descent. We choose a point in parameter space to be our truth parameters, and randomly sample a second point. From here, we compute the loss between the two samples and its gradient via Maximum Mean Discrepancy (MMD). The gradient estimate is then used to choose the next point in parameter space, and we repeat this process until the loss falls below a given threshold. We perform 10 extra iterations at the end where we double the sample size to reduce the variance of the estimator. In Figure \ref{fig:params_evolution} we show this procedure in action for two gradient descent paths. The procedure works well in general but suffers from occasional instability, especially for smaller sample sizes. This is illustrated in Figure \ref{fig:loss_alpha}, where we can see the variance in the loss estimated for different values of the $\alpha$ parameter. Convergence is still achieved when computing probabilities and gradients for only a fraction of the total discrete paths, promising scalability for a more realistic implementation of the model.

\section{Outlook and conclusions}
We have demonstrated the feasibility of a fully differentiable implementation of the key components of a nuclear deexcitation model for low energy neutrino physics, with fast and robust sampling and gradient estimation. Immediate work should be done in the stability of gradient descent, but we are confident on our approach and its scalability.

The next steps for our differentiable deexcitation simulation include:
\begin{itemize}
    \item A more complete implementation of the Hauser-Feshbach deexcitation model.
    \item Work on stability and speed of the gradient descent algorithm.
    \item Individual event moving (see \cite{bib:parton}) with higher order derivatives.

\end{itemize}
In the longer term, it would be ideal to include the neutrino-nucleus interaction physics in our model, and to integrate it in a fully differentiable simulation chain, from generation to reconstruction.

% \begin{figure}[ht!]
%     \centering{\includegraphics[width=0.6\columnwidth]{gradient_paths.png}}
%     \caption{Three deexcitation processes in the continuum.}
%     \label{fig:deex-paths}
% \end{figure}


\newpage

\providecommand{\href}[2]{#2}\begingroup\raggedright\endgroup

\end{document}